\documentclass[acmsmall,authorversion]{acmart}
\usepackage[utf8]{inputenc}
\usepackage[T1]{fontenc}
\usepackage[english]{babel}

\usepackage{graphicx}
\usepackage{xspace}
\usepackage{subcaption}

\usepackage{enumitem}
\setlist{itemsep=0mm}
\usepackage{amssymb}
\usepackage{blindtext}
\usepackage{array}
\usepackage{multicol}
\usepackage{wrapfig}

\usepackage{tikz}
\usetikzlibrary{decorations.text,backgrounds,positioning,shapes,
  shadings,shadows,arrows,decorations.markings,calc,fit,fadings,
  tikzmark
}

\usepackage[scaled=0.85]{DejaVuSansMono} 

\usepackage{natbib}
\addto\extrasenglish{%

}

\newcommand{\mirage}{MirageOS\xspace}

\newcommand{\ocaml}{OCaml\xspace}

\newcommand{\code}[1]{\texttt{#1}}

\usepackage{xcolor}
\definecolor{butter}{HTML}{C4A000}
\definecolor{orange}{HTML}{CE5C00}
\definecolor{chocolate}{HTML}{8F5902}
\definecolor{chameleon}{HTML}{4E9A06}
\definecolor{skyblue}{HTML}{204A87}
\definecolor{plum}{HTML}{5C3566}
\definecolor{scarletred}{HTML}{A40000}
\definecolor{lightalu}{HTML}{BABDB6}
\definecolor{darkalu}{HTML}{2E3436}

\newcommand{\kwstyle}{}

\usepackage{listings}

\lstMakeShortInline[basicstyle=\normalsize\ttfamily]{|}

\title{Programming Unikernels in the Large via Functor Driven Development (Experience Report)}

\author[G. Radanne]{Gabriel Radanne}
\affiliation{
  \institution{University of Freiburg}
}
\author[T. Gazagnaire]{Thomas Gazagnaire}
\affiliation{\institution{Tarides}}
\author[A. Madhavapeddy]{Anil Madhavapeddy}
\author[J. Yallop]{Jeremy Yallop}
\author[R. Mortier]{Richard Mortier}
\affiliation{\institution{University of Cambridge}}
\author[H. Mehnert]{Hannes Mehnert}
\author[M. Preston]{Mindy Preston}
\affiliation{\institution{Robur}}
\author[D. Scott]{David Scott}
\affiliation{\institution{Docker, Inc}}

\begin{document}

\begin{abstract}

  Compiling applications as unikernels allows them to be tailored to diverse
  execution environments. Dependency on a monolithic operating system is replaced with linkage against libraries that provide specific services.
  Doing so in practice has revealed a major barrier: managing
  the configuration matrix across heterogenous execution
  targets.  A realistic unikernel application depends on hundreds of libraries, each of which may place different demands on the different target
  execution platforms (e.g.,~cryptographic acceleration).

  We propose a modular approach to structuring large scale
  codebases that cleanly separates configuration, application and operating
  system logic. Our implementation is built on the \mirage unikernel
  framework, using the \ocaml language's powerful abstraction and
  metaprogramming facilities. Leveraging modules allows us to build many
  components independently, with only loose coupling through a set of standardised
  signatures.  Components can be parameterized by other components and composed.
  Our approach accounts for state, dependency ordering, and error
  management, and our usage over the years has demonstrated significant efficiency benefits
  by leveraging compiler features such as global link-time
  optimisation during the configuration process.
  We describe our application architecture and experiences via some practical applications of our approach, and 
  discuss how library development in \mirage
  can facilitate adoption in other unikernel frameworks and programming languages.

%

\end{abstract}

\maketitle
\begin{figure}[!h]
  \centering
  \input{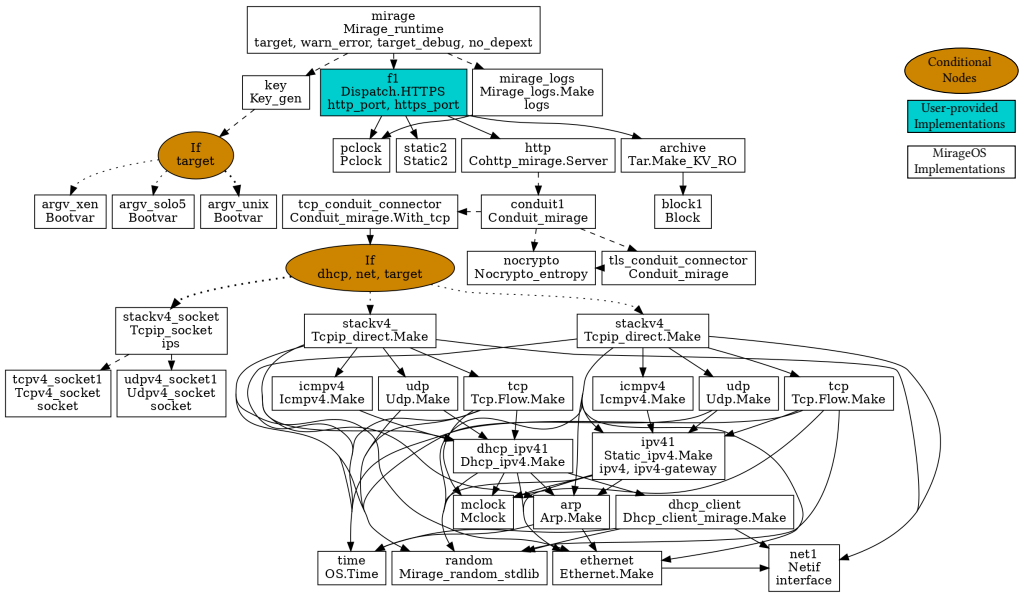}
  \vspace{-1.3em}
  \caption{Configuration graph for a \mirage web server}
  \label{teaser}
  \vspace{-1em}
\end{figure}

\section{Introduction}

A major source of complexity in modern application development is the need to run on an increasingly diverse range of platforms:
   conventional operating systems (OSs) such as Linux or Windows, 
   mobile systems such as Android or iOS,
   embedded platforms such as ARM or RISC-V microcontrollers,
   or browser-based virtual machines via compilation to JavaScript or WASM.
Designing efficient programming interfaces for such heterogenous environments is challenging
  as all have different internal models
                    and mechanisms for
                          memory management,
                          isolation,
                          I/O
                          and scheduling.

Attempts to adapt existing models (e.g.~POSIX) to these
environments has led to a lowest common denominator set of ``mini-libc'' system
libraries. These are deeply unsatisfying: one would rather generate binaries
that are specialised for a particular platform, able to make full use of its specific capabilities.
Ideally, we would
have a modular set of interfaces allowing applications to depend on
the specific functionality they need to operate on a specific physical or virtual platform.

One key step towards this goal is to use library operating
systems (libOSs) to break down monolithic kernel components into conventional
libraries that can be linked alongside application logic~\cite{nemesis,exokernel}.
When these kernel and application libraries are linked to a bootloader, the
result is a single-purpose {\em unikernel}, specialised at build time
to execute that specific application on the specific platform~\cite{unikernel}.
The specialisation has been shown to result in significant performance
and code-size improvements in the resulting artefacts~\cite{jitsu,Manco:2017:MVL:3132747.3132763}.

The last few years have seen many new unikernel frameworks written in
high-level languages.
The number of kernel libraries has grown concomitantly, resulting in a
practical challenge:
 {\em how can developers avoid the need to manually select the set of
   kernel libraries required by a target platform}?
The common approach of depending on a monolithic OS interface
layer (e.g.,~in \ocaml, the {\tt Unix} module), does not scale to the
modern heterogeneous world.

This paper describes our experiences in addressing this problem of
writing high level code that can run in heterogenous execution
environments by using \ocaml{}'s powerful abstraction facilities
within the \mirage unikernel framework.
\mirage has been developed since 2006 and has seen widespread deployment in industrial projects such as
Xen~\cite{Scott:2010:UFP:1932681.1863557,Gazagnaire:2009:OEH:1596550.1596581} and Docker.  Over the last
decade, \mirage has grown to support a highly  diverse set of target platforms including hypervisors such as Xen~\cite{xen}, KVM~\cite{kvm} and Muen~\cite{muen}, plus conventional
Unix and Windows binaries, and even experimental compilation to JavaScript and bare-metal booting on RISC-V and ARM boards.

The key challenge in maintaining these compilation targets has been to
prevent \ocaml programmers, otherwise fastidious about their use of
abstraction, from using the monolithic OS interfaces such as \texttt{Unix}
that tie an application to a single execution environment.
Instead, \mirage takes advantage of the powerful ML module system to
allow programmers to abstract over use of individual OS facilities
(e.g.,~timekeeping, networking, storage, entropy).
Rather than calling into \texttt{libc}, application code is abstracted
over the OS functionality needed using \ocaml{}'s parameterised
modules.
The \mirage compiler then supplies library implementations of the
required functionality suitable for the target platform.
These implementations range from trivial passthroughs that
invoke system calls on Unix, to complete
reimplementations of key kernel subsystems such as TCP/IP for targets
without a conventional OS such as bare-metal
embedded devices or Xen hypervisors.

This way, developers write application code that can be efficiently
compiled to any of these environments simply by making their
dependencies on system facilities explicit using parameterisation.
The resulting codebases are also highly structured (see~\autoref{teaser}
for the MirageOS webserver) and easily compiled to future deployment targets.
We dub this approach \emph{Functor Driven Development}, and make the
following experience contributions in this paper:
\begin{itemize}[leftmargin=*]

\item we describe our portable application structuring that encourages
  developers to explicitly specify OS dependencies by using \ocaml
  module constructs: structures, signatures, and functors
  (i.e.~functions over modules) (Section~\ref{sec:syslib});
 
\item we show how we make use of meta-programming techniques to
  generate the complex glue code that connects configuration, build
  and deployment of the application, using an eDSL to express
  dependencies between the application requirements and concrete
  implementations for a particular target platform
  (Section~\ref{sec:functoria}); and

\item discuss our experiences with using the OCaml module system
  at scale for operating system assembly (Section~\ref{discussion}).

\end{itemize}

%

%



\section{Structuring Applications with Functors}
\label{sec:syslib}

\ocaml modules~\citep{MacQueen84,Leroy94,Leroy95}
allow  programs to be built from smaller components.
In most languages, modules are compilation units:  simple collections of
type and value declarations in a file.
\ocaml extends such collections, called \emph{structures},
with \emph{signatures} (module types),
     \emph{functors} (functions from modules to modules)
     and functor application, to
form a small typed functional language.
Developers use this language to group, compose and selectively expose
program components (types, values, functions, and modules).
Modules are structurally typed: a module need not announce which
signatures it satisfies, and a single module can satisfy many different
signatures, which may expose or conceal module components, and
present types as concrete or abstract.
Modules may be combined using functors, which construct new modules
from existing modules passed as arguments.

\begin{figure}[!b]
  \centering
  \begin{minipage}[t]{0.55\linewidth}
\begin{lstlisting}[linerange=1-8]
module type Store = sig
  type t
  val read: t -> string -> string
end
module type Network = sig
  type t
  val listen:t -> (string -> string) -> unit
end
\end{lstlisting}
  \end{minipage}\hfill
  \begin{minipage}[t]{0.45\linewidth}
\begin{lstlisting}[linerange=10-13,firstnumber=last]
module Make (S: Store) (N: Network) = struct
  let start storage network =
    N.listen network (S.read storage)
end
\end{lstlisting}
  \end{minipage}
  \vspace{-2em}
  \caption{A modular file server.}
  \label{code:server:modular}

  \centering
  \begin{subfigure}[t]{\textwidth}
    \begin{minipage}[t]{0.33\textwidth}
\begin{lstlisting}
module Direct : sig
 (* read host filesystem *)
 include Store
 val create : string -> t
end
\end{lstlisting}
      \vspace{-1em}
    \caption{\code{Direct} implements \code{Store}.}
    \label{code:store:direct}
    \end{minipage}%
    \begin{minipage}[t]{0.33\textwidth}
\begin{lstlisting}
module Crunch : sig
 (* read compiled-in strings *)
 include Store
 val create : Crunch.t -> t
end
\end{lstlisting}
      \vspace{-1em}
    \caption{\code{Crunch} implements \code{Store}.}
    \label{code:store:crunch}
    \end{minipage}%
    \begin{minipage}[t]{0.33\textwidth}
\begin{lstlisting}
module NetStore (N: Network): sig
 (* read from network service *)
 include Store
 val create : N.t -> t
end
\end{lstlisting}
      \vspace{-1em}
    \caption{\code{NetStore} implements \code{Store}}
    \label{code:store:depot}
    \end{minipage}%
  \end{subfigure}
  \caption{Examples of store implementations}%
  \label{code:store}
  %

  \begin{subfigure}[t]{\textwidth}
    \begin{minipage}[t]{0.5\textwidth}
\begin{lstlisting}
module TCPIP : sig
  include Network
  val create : int -> t
end
\end{lstlisting}
      \vspace{-1em}
    \caption{\code{TCPIP} implements \code{Network}.}
    \label{code:network:tcp}
    \end{minipage}%
    \begin{minipage}[t]{0.5\textwidth}
\begin{lstlisting}
module HTTP (N : Network) : sig
  include Network
  val create : N.t -> t
end
\end{lstlisting}
      \vspace{-1em}
    \caption{\code{HTTP} implements \code{Network}.}
    \label{code:network:http}
    \end{minipage}%
  \end{subfigure}
  \caption{Examples of network implementations}
  \label{code:network}
\end{figure}

\autoref{code:server:modular} uses this technique to design a simple
static file server
with two module parameters:
  |S| of type |Store|, which describes how to access local files,
  and |N| of type |Network|, which describes how networking is managed.
|Store| and |Network| each expose
  a type |t|
    (representing the storage and network handles respectively)
 and a function:
    |listen| makes a callback that listens on the network handle,
    and |read| accesses the current store to read a file.
The core application logic is defined by the functor |Make| whose body contains a single function that calls
the (abstract) functions from its module parameters
|N| and |S|.

\autoref{code:store} and \autoref{code:network}
show several storage and network implementations.
\code{Direct} (\autoref{code:store:direct}),
\code{Crunch} (\autoref{code:store:crunch}) and
\code{NetStore} (\autoref{code:store:depot})
implement various kinds of \code{Store}.
As well as satisfying the signature |Store|, each implementation also
provides a |create| function with specialised arguments to take care
of device-specific initialization.
\code{Direct.read} gives access to the underlying
filesystem, the handle being the root of the filesystem in question. 
%
\code{Crunch} provides  an in-memory representation of a file-system.
It operates by turning a filesystem tree into an
\ocaml module which is then compiled and embedded in the application at configuration time.
Finally, \code{NetStore} presents an online service as an
initially-empty \code{Store}; it processes requests to add files.
\code{NetStore} requires network access and is thus a
functor parameterised by a module of type \code{Network}.

\code{TCPIP} (\autoref{code:network:tcp}) and
\code{HTTP} (\autoref{code:network:http})
implement \code{Network}.
The function \code{TCPIP.listen} uses the POSIX \code{listen} and \code{accept}
syscalls to handle incoming TCP/IP connections on the given port. It
then reads a request line and returns the result of passing it to a
callback.
The function \code{HTTP.listen} handles connections, reading a full HTTP request
when a client connects, extracting the HTTP path and passing it to the
callback.  The resulting file content is wrapped into an HTTP response
by adding the correct headers, before being returned to the client
connection. Note that this implementation depends on another network
stack to simply read request and response contents without
interpretation. We can use this to implement HTTP over TCP/IP or over
TLS to get HTTPS.

Each of these modules can be used to satisfy the application's functor
allowing our simple static fileserver to target a very wide range of
deployment platforms.  In each implementation, the type |t| represents
wildly different states but, as |t| is abstract, \ocaml ensures that
details of the type's implementation are never used in the body of the
|Make| functor in \autoref{code:server:modular}.

\subsection{Standardized Signatures}
\label{sec:syslib:sig}

Our example application consists of two major pieces of external
functionality: file system access and networking.
MirageOS separates these two domains from the usual monolithic
\texttt{Unix} module by defining independent module signatures, which
are then implemented by several modules.
This modular approach has two advantages: it avoids a dependency on a
monolithic OS kernel, and it disaggregates functionality into smaller
module signatures that can be separately implemented by experts in
each domain.
File system experts can contribute implementations of the |Store|
signature, and network developers can write |Network| implementations.
The signature approach also makes dependencies between different
domains explicit; for example, the |NetStore| implementation interacts
with both |Network| and |Store|.

This strong isolation of concerns has proven essential in growing the
MirageOS ecosystem.
An operating system contains many pieces pertaining to very different
domains.  \mirage contains libraries ranging from bare-metal drivers
to TLS implementations, including high-level HTTP servers.
Contributors' knowledge in a given domain can be applied to build
additional implementations that will fit into the overall ecosystem,
without getting overwhelmed by the enormity of the full clean-slate
operating system stack.

Having implementations bundled as modules with a common interface is
also beneficial for testing purposes.
Complex components can be tested in isolation and often without
requiring a physical environment.
Tests can be expressed as functors over the signatures to test,
allowing us to stress the implementation in a virtual environment
convenient for local use (e.g.~a fake networking bridge).
We also use this approach to test the applications themselves, which
are also parameterised by their module dependencies.
We have combined this parameterised testing approach with property
testing~\citep{DBLP:conf/icfp/ClaessenH00} and
fuzzing~\citep{crowbar,afl} in various implementations.

\subsection{State and Initialization}
\label{sec:sys:state}

All the functors and modules in the previous sections are \textit{pure}: they
do not produce side effects when applied to other modules.
Applying a functor creates a new module built from its parameters, but
does not perform initialization or modify state.
This is convenient for two reasons.  First, modules might share an
interface for most of their operations except for the initialization
code.  For example, in our store implementations
(\autoref{code:store}) the type of |create| varies with each
implementation, but besides |create| the implementations all simply
implement the |Store| signature.
By separating initialisation functions from the rest of the
operations, we ensure that the core application, such as the |Make|
module in \autoref{code:server:modular}, can be used with a large
variety of implementations.
Second, purity maximises implementation sharing without mixing up
state.
For example, in the |NetStore| module we might share the same
|Network| implementation for both the online repository and to serve
files.
However, although the implementations are shared, the network handle
itself is not, ensuring we don't accidentally couple the two
otherwise-separate components.

Although initialization code might be different for each module, there
are some regular patterns that inform our signature design.
In the main function of our fileserver in
\autoref{code:server:modular} we require a store and a network handle,
which correspond to the two arguments of the functor.
This pattern is both common and expected: for functors, the
initialization function typically requires the results of the
initialization of each module arguments.  This property holds in all
the functors we have presented so far.

\subsection{Reporting Errors}
\label{sec:syslib:error}

A modular system that allows for many implementations must also
provide some mechanism for reporting errors. This error information
must be simultaneously fine-grained enough for the developer to
determine the appropriate recovery or failure mechanism, and coarse
enough for different implementations to provide reasonable information
in each possible failure case.

In \mirage, we eschew the use of exceptions in favour of a more
explicit approach using the standard |result| type and \ocaml's
\emph{polymorphic variants}~\cite{DBLP:conf/aplas/Garrigue01}.
The |result| type is a binary sum: a value of type |result| is either
a ``success'' value |Ok v| or an ``error'' value |Error err|.  Result
also comes with monadic operations for chaining computations that can
fail.
%
%
OCaml's structurally typed polymorphic variants are distinguished by a
leading backtick \code{`} for each constructor: for instance,
|`Unknown_file s| has the type |[> `Unknown_file of string]|.
Using structural typing makes it possible to combine multiple error
types.
For example, if |store_error| and |network_error| are polymorphic
variant types, then \lstinline[basicstyle=\normalsize\ttfamily]{[> store_error | network_error]}
denotes the combination: any value of
either |store_error| or |network_error| is also a member of this type.

\autoref{code:store:error} extends the |Store| signature
to use these extensible error types.
The revised \code{Store} signature exposes a type \code{error},
consisting of general errors expected to be encountered by any |Store|
implementation, along with a pretty printer\citep{FormatUnraveled}
that builds a human-readable representation of an error
(\autoref{line:store:pp_error}).
By making the \code{error} type |private|~\citep{10.1007/11924661_3},
we allow the implementation to provide a richer error type, as long as
it contains at least the specified elements.
Module type signatures with functions that may return an
error use the \code{result} type to return either the result of a successful call or the
relevant error information.
For example, \code{Store} uses the \code{error} type together with \code{result} to provide
structured error reporting for the |read| function (\autoref{line:read:error}).

This design has several appealing features.
First, errors are extensible: individual implementations of
\code{Store} can extend the \code{error} type with
implementation-specific errors,
Second, error checking is compositional: error types from multiple
modules can be combined, and users can leverage the monadic API of the
|result| type to chain computations.
Third, error-checking is typed: \ocaml{}'s type system ensures that
clients that abstract over \code{Store} signature can only match on
errors (such as \texttt{Unknown\_file}) exposed by the signature
(although the pretty printer can always be called to log messages
about other errors).

\begin{figure}[!bt]
  \begin{minipage}[b]{0.48\linewidth}
\begin{lstlisting}
module type Store = sig
  type error = private [> `Unknown_file of string]
  val pp_error: Format.formatter -> error -> unit(*@\label{line:store:pp_error}*)

  type t
  val read:
    t -> string -> (string, error) result(*@\label{line:read:error}*)
end
\end{lstlisting}
    \vspace{-1em}
    \caption{\code{Store} extended with modular error handling.}
    \label{code:store:error}
  \end{minipage}
  \hfill
  \begin{minipage}[b]{0.47\linewidth}
\begin{lstlisting}
module Store = Crunch
module Network = Http (TCPIP)(*@\label{line:funapp:a}*)
module MyServer = Server.Make (Store) (Network) (*@ \label{line:funapp:b}*)

let () =
  let crunch = CrunchModule.data in
  let store = Crunch.create crunch in
  let tcpip = TCPIP.create 80 in
  let network = Network.create tcpip in (*@\label{line:funapp:c}*)
  let server = MyServer.start store network in(*@\label{line:funapp:d}*)
  run server
\end{lstlisting}
    \vspace{-1em}
    \caption{Bringing it all together}
    \label{code:server:functorapp}
  \end{minipage}
\end{figure}

\subsection{Gluing Modules Together}

\autoref{sec:syslib} described a modular file server and showcased
several implementation for its sub-components. The flexibility of the
modular approach allows us to assemble our application in a LEGO
fashion by plugging modules together.
\autoref{code:server:functorapp} combines the various components to
create a self-contained file server that can be used in a POSIX
environment. We use the |Crunch| module along with the |HTTP| functor
applied to |TCPIP|.  This results in two functor applications
(Lines~\ref{line:funapp:a}-\ref{line:funapp:b}). We then need to
initialize the various elements of our fileserver and launch it
(Lines~\ref{line:funapp:c}-\ref{line:funapp:d}).  Note how the
initialisation code closely reflects the structure of the functor
instantiation code, thanks to the regular pattern noted in
\autoref{sec:sys:state}.

Although it is straightforward, this code is not completely
satisfactory to write by hand.
Firstly, the code is repetitive: the structure of the functor
applications and the state initialization is the same in each case
(For example, the function applications in lines~\ref{line:funapp:c}
and \ref{line:funapp:d} mirror the functor applications in
lines~\ref{line:funapp:a} and~\ref{line:funapp:b}.)
Furthermore, the code must be modified by hand each time we change a
component of our application.
(For example, using |Direct| in place of of |Crunch| would require
changing both the functor applications and the initialization by hand.)
Finally, while the code in this toy example is rather simple, its
complexity rapidly increases in a realistic application.
(For example, the unikernel that runs the \mirage website contains
more than 70 modules and a functor application depth of up to 10 for
the devices it uses.)
To handle such a rich ecosystem, we need better tooling.


\section{Functoria: A Tool to Glue Modules and Signatures Together}
\label{sec:functoria}

Building executable applications from functor-heavy libraries involves
significant boilerplate.
\ocaml's module language is much less flexible than its expression
language: it does not support conditionals or more complex dependency
requirements.
This section presents a tool {\em functoria} and its DSL that acts as
the glue language between the module and expression portions of the
MirageOS application, allowing us to overcome these limitations.

The high-level goal of functoria is to automatically configure and
build modular applications, such as the file server presented in
\autoref{sec:syslib}, across the full variety of MirageOS backends.
Functoria provides a CLI interface which takes arguments pertaining to
the application to explicitly configure each of the constituent
modules:

\begin{itemize}[leftmargin=*]
\item |functoria configure --store direct --fs /my/files -p 42|\\
  configures the application to serve ``/my/files'' over a socket on port 42.

\item |functoria configure --store crunch --fs /my/files -p 80|\\
  configures the application to create an HTTP file server serving the crunched
  files in ``/my/files'' on port 80.  (The logic that interprets |80| to build
  an HTTP server is described below.)
\end{itemize}

The configuration process generates a file |main.ml| that applies all the
application functors with concrete implementations, and also invokes the
device initialisation code with the supplied configuration parameters.
All the programmer has to do is to install any OCaml dependencies and invoke |make|
to generate an executable unikernel from |main.ml|.

\subsection{Configuring applications with functoria}

Functoria relies on a configuration language that acts as a well-typed enforcer
of the {\em structure} of the application (expressed by the programmer by
functorising across its dependencies) and the {\em implementation} of those
dependencies (expressed during the configuration process).

\begin{figure}[!t]
  \centering
  \begin{minipage}[t]{0.47\linewidth}
\begin{lstlisting}
type 'a typ
val (@->): 'a typ -> 'b typ -> ('a -> 'b) typ

type 'a impl
val ($): ('a -> 'b) impl -> 'a impl -> 'b impl

val foreign: -> string -> 'a typ -> 'a impl
\end{lstlisting}
    \vspace{-1em}
\caption{Library to describe modules and functors}
\label{functoria:api}
    \centering
\begin{lstlisting}[firstline=3]
type store
val store : store typ

type network
val network : network typ

val direct : string -> store impl
val crunch : string -> store impl
val netstore : (network -> store) impl
val tcpip : network impl
val http  : (network -> network) impl
\end{lstlisting}
    \vspace{-1em}
    \caption{Devices combinators for the file server}
    \label{code:server:devices}
  \end{minipage}\hfill%
  \begin{minipage}[t]{0.47\linewidth}
    \centering
    \begin{lstlisting}
let make_server =
  foreign
    "Server_modular.Make"
    (store @-> network @-> job)
    
let my_server =
  make_server $
  direct "data/" $
  (http $ tcpip)

let () = register "filesrv" [ my_server ]
\end{lstlisting}
    \vspace{-1em}
  \caption{A \code{config.ml} file for the file server}
  \label{code:server:configfile}

  \vspace{1em}
  
  \includegraphics[width=.95\linewidth]{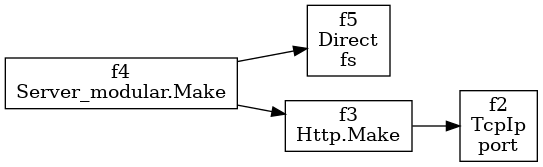}
  \vspace{-1em}
  \caption{Configuration graph for the file server}
  \label{fig:server:configdot}
\end{minipage}

\end{figure}

\autoref{functoria:api} shows functoria's high-level operations for
describing functors.
A value of type \code{typ} represents a module type such as
\code{Store} or \code{Network}.
The \code{@->} operation builds a functor type from the types of its
parameter and result: \code{store @-> network @-> job} represents the
type of a functor that takes module arguments of type \code{store} and
\code{network} and builds a module of type \code{job} representing the
final unikernel.
A value of type \code{impl} represents a module implementation.  There
is one operation, |$|, that corresponds to module application.
The \code{foreign} function materializes a named module (i.e.~creates
a value of type \code{impl}) given its name and type.

Functoria also exposes particular values of type \code{typ} and
\code{impl}, for the signatures and modules available in \mirage
(\autoref{code:server:devices}).
For example, \code{store} (of type \code{store typ}) corresponds to
the \code{Store} signature (\autoref{code:server:modular}), and
\code{direct} (of type \code{store impl}) corresponds to the
\code{Direct} implementation of type \code{Store}
(\autoref{code:store:direct}).
In each case the type index serves as a witness to ensure signature
compatibility. These \code{typ} and \code{impl} values can be used for
reflection (e.g.~to list all the available implementations available
for a given signature) as well as for composing functors to build
devices.

Functoria also allows describing the various metadata associated with
a module such as the packages it requires from OPAM (the OCaml package
manager).
Indeed, modules described by the configuration do not have to be
immediately available in the current environment, but can be present
in external libraries. The functoria tool will use OPAM to install all
the required dependencies.

To configure a unikernel using these operations, the programmer
creates a file |config.ml| that specifies how to combine the various
module implementations.
\autoref{code:server:configfile} shows an example that corresponds to
the handwritten code of \autoref{code:server:functorapp}.
The value |make_server| represents a functor |"Server_modular.Make"|
with two parameters.
The value |my_server| represents an application of that functor to two
arguments: the module |direct|, and the result of applying the functor
|http| to the server |tcpip|.
Finally, the {\tt register} function specifies and names the main
module of the application.

Based on this code, Functoria will derive a graph that describes the
structure of the application (\autoref{fig:server:configdot}).
This graph is used to synthethise everything related to the
application: dependencies, initialisation and module code,
documentation, package manager invocations, and so on.

\subsection{Parametrized applications}
\label{sec:paramapps}

\begin{figure}[!t]
  \begin{minipage}[t]{0.54\linewidth}

  \begin{lstlisting}
module Key : sig
  type 'a k
  type +'a v
  val create: string -> 'a Arg.t -> 'a k
  val value: 'a k -> 'a v
  val pure: 'a -> 'a v
  val ($): ('a -> 'b) v -> 'a v -> 'b v
end

val if_: bool Key.v -> 'a impl -> 'a impl -> 'a impl
val match_: 'b Key.v -> ('b * 'a impl) list ->  'a impl
\end{lstlisting}
      \vspace{-1em}
\caption{Keys for parameterised applications.}
\label{functoria:api:keys}
  \end{minipage}\hfill%
  \begin{minipage}[t]{0.45\linewidth}
\begin{lstlisting}
let default_store dir : store impl =
  let i = Key.Arg.info
      ~doc:"Choose store" ["store"] in
  let arg = Key.Arg.(opt string "crunch"
      ~stage:`Configure i) in
  let key = Key.create "store" arg in

  match_ (Key.value key) [
    "crunch", crunch dir;
    "direct", direct;
  ] ~default:(crunch dir)
\end{lstlisting}
      \vspace{-1em}
\caption{Using keys in a configuration pass}
\label{functoria:api:usage}

\end{minipage}
\end{figure}

The applications we have seen so far are very static:
changing one of the modules requires rewriting either the
code or the configuration.
To provide the kind of flexibility needed in \mirage applications,
Functoria adds an additional ingredient: \emph{keys}.
The |Key| module (\autoref{functoria:api:keys}) represents CLI
arguments that can be used during configuration to determine which
implementations to use in the generated code.

\autoref{functoria:api:usage} gives a new implementation of |Store|
that supports selecting the storage mechanism at build time.  The
|default_store| value exposes the option |--store| to the command line
and uses it to choose between the modules |Direct| or |Crunch|.  The
|Key.create| function declares a new key and the |Arg| module
describes the CLI arguments (in this example, a simple enumeration).
Finally, the |match_| function chooses an implementation based on the
CLI key selection.

From a user perspective, this allows functoria to provide some extremely
useful features for development. 
The user can choose between a filesystem or a built-in crunch store directly
from the command line (e.g. by running |functoria configure --store crunch|).
Functoria also generates the documentation of the application
that describes all its keys, both as a Unix manual page and via the CLI:
\begin{lstlisting}
$ functoria describe
Name       filesrv
Build-dir  .
Keys       store=crunch (default)
\end{lstlisting}

The |--store| key is only used during configuration; the
|match_| combinator can only swap modules at configuration time.
We use the |~stage:`Configure| argument to constrain this key
to work at configure time. However, it is also possible to use
keys dynamically at runtime. To demonstrate this, we can a new key
to our file server to determine the port to listen to.
\begin{lstlisting}
let port =
  let arg = Key.Arg.(opt int 80 (info ["p";"port"]))
  in Key.create "port" arg
\end{lstlisting}


We then use this key in the initialization code of the |TCPIP| module:
\begin{lstlisting}
let network = TCPIP.create (Key_gen.port ())
\end{lstlisting}

The |--port| option can now be provided during both configuration and
application startup. If the option is present during
configuration, the value will be persisted and used as a default value during
startup.  MirageOS backends can supply more specific implementations for
dynamic key lookup at runtime (for instance, via bootloader arguments,
browser APIs, or conventional Unix environment variables).

In the examples so far, we have used keys in a ``direct'' manner: either by
using their value directly for configuration (in the case of |--store|)
or by passing the value off to the underlying application (for |--port|).
We can also use keys for computations. For example, we define |default_network|
which uses the |HTTP| functor if the port is 80 or 8080, but uses the
normal |TCPIP| device otherwise.
We use the fact that keys are split into two types: |Key.k|, which can be passed
down to the runtime, and |Key.v|, which cannot be serialized but can be used in
computations. 
We can use |Key.value| to obtain the value associated with |port|, and then
apply |Key.pure| to our predicate to create a value that is not associated
with a key. |$| allows us to apply the previous value to |port|.
We can then use the resulting boolean value with |if_| to switch
beetwen implementations.
\begin{lstlisting}
let default_network : network impl =
  let is_http = Key.(pure (fun x -> x = 80 || x = 8080) $ value port) in
  if_ is_http (http $ tcpip) tcpip
\end{lstlisting}

|Key.value| equiped with |pure| and |$| (also often named |app|) forms 
an \emph{applicative functor}%
\footnote{In the categorical sense. Not to be confused with ML functors!}. The full library also provides other common applicative
operators such as |map|.

\begin{wrapfigure}{r}{0.4\textwidth}
\centering
\vspace{-1em}
\begin{lstlisting}[firstline=1]
val impl: 'a configurable -> 'a impl

class type ['ty] configurable = object
  method ty: 'ty typ
  method name: string
  method module_name: string
  method keys: key list
  method connect:
    Info.t -> string ->
    string list -> string

  method packages: package list Key.value
  method configure: Info.t -> unit
  method build: Info.t -> unit
  method clean: Info.t -> unit
end
\end{lstlisting}
\vspace{-1em}
  \caption{API for configurable devices}
  \label{functoria:api:configurable}
\vspace{-1em}
\end{wrapfigure}

\subsection{Sharing and configuring devices}
\label{sec:sharing}

The |foreign| function is a specialised version of {\em configurable devices}.
Configurable devices have an interface
that describes the metadata provided by |foreign| modules (type, names, package
descriptions and keys) and also the complete lifetime of a device: how to configure,
build and detach it (\autoref{functoria:api:configurable}).
The |connect| method specifies how to initialize the device---via a simple call to
|start| in the case of |foreign| devices, but arbitrary initialization code in general for
more complex cases.
Once a configurable device has been defined, it can be encapsulated as an
implementation via |impl|.


The OCaml object system proves useful here.
The definition of configurable devices using OCaml classes makes it
possible to easily define classes of devices that are more specialised
for a particular purpose.
For example, we could define a system where every device, once
initialized, must add itself to a global list of devices.
This can be encapsulated in functoria by providing a new function that
generates the appropriate initialization code and used instead of
|foreign|.


Various devices can sometimes have common dependencies. For example, a
network device can be used both by HTTP devices and DHCP devices.
However, it can't be assumed that devices are reentrant: many
drivers for network connections should not initialize twice.

In functoria, devices are identified by both a module, which indicates
their implementation, and a name, which defines
their state.  Functoria uses this name to decide which devices should be merged.
If two devices have the same names, keys and--in the case
of functors--are applied to the same arguments, they are considered
equal. Equal devices share their state and their code.
To force two devices to \emph{not} be shared, it is sufficient to give
them different names.


\subsection{Building portable and flexible applications}

\begin{figure}[!htb]
  \centering
  \begin{minipage}[b]{0.48\linewidth}
    \begin{lstlisting}
let make_server =
  foreign "Server_modular.Make"
    (store @->
     network @-> 
     job)

let my_server =
  make_server
  $ default_store "data/"
  $ default_network
\end{lstlisting}
    \vspace{-1em}
  \caption{\code{config.ml} file for the file server application}
  \label{code:server:configfile:full}
\end{minipage}\hfill%
  \begin{minipage}[b]{0.45\linewidth}
    \includegraphics[width=.9\linewidth]{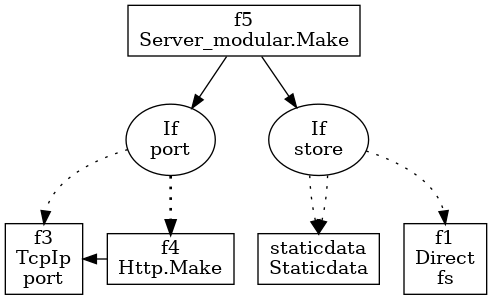}
    \vspace{-1em}
    \caption{Configuration graph of the file server.}
    \label{fig:server:configdot:full}
  \end{minipage}

\end{figure}

We have made our example application more flexible than a typical
monolithic Unix application, and are now able to change all the aspects of
our file server simply by providing command line options.
Our final configuration file, however, is barely more complex than it was
at the beginning:
\autoref{code:server:configfile:full}.
Thanks to the interfaces provided by functoria, MirageOS implementors can
provide combinators to make their devices easily usable in application configurations.
The cost of this flexibility, of course, is a multiplication of command line options and devices.
Functoria presents the configuration graph of the application in several formats
to make it easier to reason about its modular structure. The graph
for our final file server (\autoref{fig:server:configdot:full}) shows
configurable devices (rectangular nodes with a name and keys) and conditional configuration on
keys (round nodes).

\section{Discussion and Related Work}
\label{discussion}

\paragraph{Growth of the OCaml libOS ecosystem}

We have successfully used functoria as the core configuration language in MirageOS for the past three
years. During that time it has scaled to manage the ever-expanding set of OS libraries written
in pure OCaml to replace the original unsafe C versions. Functoria has been used to create many unikernel
applications such as the self-hosted website whose configuration graph is rendered in~\autoref{teaser}.
The original vision of MirageOS was to provide a {\em complete} reimplementation of
OS functionality in a type-safe language, and today the set of functoria module signatures in \autoref{ecosystem} show how far we have come in achieving this goal.

The {\texttt mirage} organisation on GitHub hosts over 100 repositories of independent OS libraries.
MirageOS supports a variety of deployment targets and the examples in the ``skeleton'' repository compile
to all of them.
Some of the available MirageOS targets are:
\begin{itemize}
  \item {\tt unix}: maps filesystem and networking through to the Unix |libc| interfaces, resulting in a standard Unix application. This mode is useful during development of higher-level logic.
  \item {\tt xen}: eliminates the dependency on a general-purpose OS and constructs a standalone kernel that boots on the Xen hypervisor. This requires a full device driver stack written in OCaml (from DHCP to TCP/IP to HTTP to TLS) that are all supported by functoria.
  \item {\tt hvt}, {\tt virtio}, {\tt muen} and {\tt genode}: these use the Solo5 hypervisor~\cite{solo5} to run under KVM or directly on more specialised operating systems such as Meun or Genode.  They also require a complete OCaml device stack instead of relying on an underlying OS.
  \item {\tt qubes}: extends the Xen compilation target with extra device drivers to work on the QubesOS secure desktop Linux distribution, for example to firewall applications from each other.
\end{itemize}

There are also more experimental targets that link directly with embedded system bootlayers to run directly on open-source ARM or RISC-V hardware~\cite{shakti}, providing a path to building highly secure and efficient IoT infrastructure. Note that all targets do not need to support all the possible device drivers---a Unix backend can only provide network sockets and not support direct Ethernet device signatures that are exposed by the Xen backend for example.

\begin{table}[!b]
\centering
\renewcommand*\arraystretch{1.2}
\newcolumntype{Y}[1]{>{\raggedright\arraybackslash}p{#1}}
\begin{tabular}{l|Y{4cm}|p{6.5cm}}
  \bf{Module type}
  &\bf{Implementations}
  &\bf{Comments}\\\hline
  
  |Mirage_kv.RO|
  & |Crunch|, |Kv_Mem|, |Kv_unix|, |Mirage_tar|, |XenStore|, |Irmin|, Filesystems
  & Read-only key-value stores allow to pass down immutable data to the unikernels
    such as webpages, certificates, etc.
    Arbitrary filesystems can also be made into key-value stores.\\
  |Mirage_kv.RW|
  & |Wodan|
  & Read-write key-value stores such as a pure OCaml store designed to run on SSDs.\\
  |Mirage_fs.S|&|Fat|, |Git|, |Fs_Mem|, |Fs_unix|
  & Filesystem implementations.\\
  |Mirage_net.S|
  & tuntap, vmnet, rawlink
  & Send and receive network packets. \\
  |ARP|, |IP|, |UDP|, |TCP|
  & |IPV4|, |IPV6|, |Qubesdb_IP|, |Udp|, |Updv4_socket|,
    |Tcp|, |Tcpv4_socket|, \dots
  & Low-level implementations of Internet and Transport Protocols. Usually has two implementations: a complete reimplementation and one that delegates to the underlying OS. \\
  |STACK|
  & |Direct|, |Socket|, |Qubes|, |Static_IP|, |With_DHCP|
  & Network stacks encapsulated for convenient usage. The stacks usually provide
    keys to customize its usage at configure and run time.
  \\
  |RANDOM|
  & |Stdlib|, |Nocrypto|, |Test|
  & Random sources, either for normal or cryptographic purposes.\\
  |HTTP|
  & |Cohttp|, |Httpaf|
  & HTTP servers implemented in term of an underlying |STACK|.\\
  |FLOW|
  & |Conduit.With_tcp|, |Conduit.With_tls|
  & A generic abstraction for network flows that can be used
    with or without encryption.\\
  |DNS|, |DHCP|, |SYSLOG|
  & |Dns|, |Unix|, |Charrua_unix|, |Charrua|,
    |Syslog.Tcp|, |Syslog.Udp|, |Syslog.Tls|
  & Protocols for various applications such as DNS, DHCP or Syslogs
    implemented in terms of an underlying |STACK| or |FLOW|.
  \\
  &|Jitsu|, |Irmin|, \dots
  &High-level APIs that can provide extra functionalities. For instance,
    Jitsu~\cite{jitsu} can spawn new VMs on-demand.
  \end{tabular}
  \caption{The MirageOS module ecosystem available on the opam package manager}
  \label{ecosystem}
\end{table}

\paragraph{Expressivity of Functoria}
Our approach relies heavily on the \ocaml module language to succeed,
and functoria provides a partial embedding of the module system
in the expression languages.
Surprisingly, although modules have much more expressive
type systems than our embedding supports, we found our subset
sufficient for our organisational use.

Our observation is that when modules are used as a large scale organisation tool,
it is generally to reduce the need for tightly coupled source codebases.
This means converging towards a set of standardised signatures and avoiding
subtyping hierarchies. The structural aspects of OCaml modules, while still
useful, can then be emulated by nominal encodings and a use of phantom
type parameters.

It is worth noting that OCaml significantly extends
beyond the original roots of Standard ML.
Features in OCaml such as applicative functors, Modula-2 style separate
compilation and polymorphic variants
have been essential when working with such
a large number of modules.
Examination of our use of these features in a large
library such as our TCP/IP stack {\em vs} a more traditional ML implementation
in the FoxNet~\cite{foxnet} project is something we plan to examine
to assess these extensions more closely.

\paragraph{Applicative vs. Generative}
\label{sec:discussion:applicative}

Since the functors used in \mirage are pure
(see \autoref{sec:sys:state}), it is desirable to share the results of
their applications as much as possible.
\ocaml's \emph{applicative} functors~\cite{Leroy94,Leroy95} are a significant
help here: 
in \ocaml, if module $M$ is equal to $N$ then the types provided by
$F(M)$ and $F(N)$ are compatible.
In contrast, functors in Standard ML~\cite{MacQueen84} are \emph{generative}, not
applicative: any types in the applications $F(M)$ and $F(N)$ are
incompatible, even if $M$ and $N$ are equal.

In \mirage, devices are considered different only if their states or
their dependencies are different (\autoref{sec:sharing}), and
so \ocaml's applicative functors are the correct default. 
However, generative functors, which \ocaml also supports, are
occasionally useful. For instance, the \mirage logging system
relies on generative functors to create a new logging implementation
with each instantiation.
Functoria supports impure functors by
generating fresh device names for each functor application, which
prevents sharing.

\paragraph{Alternative module languages}

The constructs used in our approach can be found in module languages
different from ML. Backpack~\cite{DBLP:conf/popl/KilpatrickDJM14} introduces
a ``linking calculus'' for Haskell modules that supports
features such as abstract signatures, separate compilation
and sharing that are necessary for our approach.
Scala's class calculus also supports a rich modularity toolset
that covers most of our usecases via abstract classes and generics.
MixML~\cite{DBLP:conf/icfp/DreyerR08} introduces
structures that can be partially left abstract and filled later.
This provides all the advantages of ML modules, including genericity, encapsulation and separate compilation but also support recursive modules
which could be used for interdependent devices.




\section{Conclusion}

We have presented functor-driven development, an application architecture
that leverages \ocaml modules to structure application logic in a highly portable 
form that can be compiled across a variety of heterogenous targets.

Our implementation of the MirageOS unikernel framework has allowed us to
successfully scale our ecosystem to hundreds of OCaml libraries.
These libraries are packages which themselves contain thousands of OCaml modules.
Overall we have millions of lines of modular and reusable OCaml code that provides
clean-slate implementations of OS components -- everything from device drivers
to Internet protocols -- that can be deployed on a large (and increasing) array
of execution targets.




\bibliography{biblio}


\end{document}
